\documentclass[11pt,twoside,a4paper]{article}
\usepackage{epsf}
\newdimen\captwidth   %  Comprenez: largeur de la caption
\newdimen\figwidth   %  Comprenez: largeur de la figure

\newcommand{\heading}[1]{
        \vspace*{15mm}
        {
                \Large
                \begin{center} 
                        {\bf{#1}} 
                \end{center}
        }
}
\renewcommand{\author}[3]{
        \vspace{5mm}  
        \begin{center}
                {\normalsize \rm #1}\\    % first line of authors' names
                {\normalsize \it #2}\\    % first address
                {\normalsize \it #3}\\    % second address
                \vspace{1cm}
        \end{center}
}
\newcommand{\acknowledgements}[1]{
        \vspace{7mm} 
        \noindent 
        {\normalsize \bf Acknowledgements.\,} 
        {\normalsize #1}
}

\def\micron{\hbox{$\,\mu {\rm m}\,$}}

\def\zu{\rm\,}     % units in formulae are roman typed
\begin{document}
\heading{The case for a bolometric millimetre camera\\ 
at the IRAM 30m telescope}
\author{F.--X. D\'esert $^1$ \& A. Benoit $^2$}
    {$^1$ Laboratoire d'Astrophysique, Observatoire de Grenoble BP 53,\\ 
    414 rue de la piscine, F--38041 Grenoble Cedex 9  France\\
    e-mail: Francois-Xavier.Desert@obs.ujf-grenoble.fr}
    {$^2$ CRTBT--CNRS, 25 ave. des Martyrs, BP 166, 38042 Grenoble Cedex 9
    France\\
    e-mail: benoit@polycnrs-gre.fr 
             }
\begin{abstract}
  We describe here the important astrophysical results that could be
  obtained by using large format (say $32\times32$) bolometric
  detectors at 1 and 2~mm with the IRAM 30m telescope: having a
  confusion--limited 1~mm extragalactic survey containing a large fraction
  of high redshift objects, mapping star formation regions in our
  galaxy at 1~mm, and mapping the Sunyaev-Zeldovich effect at 2~mm in tens of
  high-redshift clusters.  We also show a first optical implementation and
  the key points of this project.
\end{abstract}

\section{Scientific goal}
\subsection{The 1~mm source counts}
Franceschini et al. (1994, see Burigana et al. 1997), Guiderdoni et
al. (1998), Blain et al. (1998a, 1998b) have given number counts
estimate at the sub--mJy level for a wavelength of 1.3mm, for various
galaxy evolution models. It seems that an episode of high rate of star
formation is required at high redshift to explain both 
\begin{itemize}
\item the submillimetre background observed with FIRAS on COBE (Puget
  et al. 1996, Guiderdoni et al. 1997, Lagache et al. 1998, Fixsen et
  al. 1998) and with DIRBE on COBE (Hauser et al, 1998)
\item and the deep surveys with SCUBA at 850 and 450\micron on the
  JCMT (Hughes et al. 1998, Eales et al. 1998).
\end{itemize}

We estimate that one can expect to typically observe 1 galaxy per
arcmin$^2$ above a flux of 1~mJy at 1.2~mm. This corresponds to one
source per 30 diffraction beams. So a deep survey with the IRAM 30m
should aim at confusion--limited maps with a noise per beam of
0.2~mJy. With an estimated sensitivity of 50~mJy.s$^{1/2}$ (see
below), this means that the camera field must be observed for 13
hours, to reach that level.  In 100~hours of integration, the number
of detected sources with flux above 1~mJy (at the $5\sigma$ level) can
be expected to be about 70. That would be a major breakthrough in
order to study the statistics of this population, even allowing for a
factor 2 uncertainty in these numbers.  If SCUBA is already finding
this population, why should we try to do this in the atmospheric
window at 1.2~mm? The answer lies in the now famous positive
K--correction that happens for high redshift objects. If SCUBA has a
rather strong redshift selection around 3, one can expect a deep 1.2~mm
survey to be biased towards redshift 5 objects.  Hence, we would probe
the evolution of the Universe at large redshifts, for which we know
next to nothing.  The large collecting area and high angular
resolutions of the IRAM 30m telescope would give us a substantial
advantage in the search of primeval galaxies. At these wavelengths,
the galactic cirrus contamination is much less than in the
submillimetre domain, because high redshift objects look colder than
the high latitude cirrus clouds.

Millimetre interferometers cannot achieve this mapping speed because
their field of view is much smaller. Competition with the future LSA/MMA for
surveys has to be carefully studied in this research area.

Surveys at 2.1~mm could be quite important as well; see a first BIMA
attempt by Wilner \& Wright (1997). The confusion limit would be reached at
0.5~mJy (0.4 galaxies per arcmin$^2$) in probably less time than at
1.2~mm ($5\sigma$ in a few hours). But this is very much dependent
on the assumptions about the very high redshift Universe (z between 4
and 10).

Blank sky surveys should be done in areas where many complementary
data have been accumulated. Obviously the HDF, CFRS and deep radio
survey fields are prime targets. Mapping fields around clusters seems
also a very powerful technique to observe the high redshift Universe,
as done with SCUBA by Smail et al. (1997).

\subsection{Mapping star formation regions}
%The gain in mapping speed could provide much information on the cold
%clouds at the origin of the star formation in our Galaxy and in nearby
%ones. This point is mentionned here but not developped further.

The gain in mapping speed will provide much more information on the cold 
clouds at the origin of the star formation in our galaxy and nearby ones
but also it will 
allow to probe the evolution of the circumstellar material around 
single and multiple young stars.

This is particularly true for some crucial subjects which are today strongly 
limited by the sensitivity of current bolometer arrays.

Among them, one can present here a few major topics:

\begin{itemize}

\item {\bf Determination of the initial mass function in nearby star-forming
region}. Today many of these studies are performed in the main isotopes of CO
in J=2-1 or J=1-0 lines because they are easy to detect. In complement to CO 
studies, deep and large surveys of the optically thin emission of the dust
would strongly improve our knowledge of the clump distribution in cloud 
interiors (Motte et al. 1998). 

\item {\bf Sensitive mapping at moderate resolution ($\sim 10''$) of the 
proto-stars (Class 0 and Class I objects)}. These mappings are fundamental 
because they allow the determination of the total amount of mass 
surrounding protostars, contrarily 
to mm interferometers wich resolve out the extended envelope surrounding 
these objecs (Gueth et al., 1997).
%Moreover, multifrequency observations (1mm and 2mm) at angular resolution of 
%same order provide an estimate of the spectral index therefore of the dust 
%properties. Today, the determination of spectral index of the dust emission 
%is strongly limited by the calibration uncertainties of the observations 
%which are performed at different resolution on different telescopes. 

\item { \bf Pre-Main-Sequence Stars (Class II objects)}
In the more evolve stage of the TTauri phase, most of the envelope has 
disapeared and the material is in the form of a Keplerian disk 
(e.g. DM Tau, Guilloteau \& Dutrey 1998) 
which remains unresolved by single-dish telescopes. However, sensitive 
observations of such objects would help to probe the amount of mass in the
outer part of the disk and the extended envelope (if any) where the dust 
escapes the detection threshold of current mm arrays.    

\item{\bf Young Stars} 
Finally, we have now several examples of young stars having dusty disks
(similar to the Beta Pic disk). Many new objects have been recently detected 
and mapped at 0.8mm wavelengths with SCUBA around Solar-type and 
Vega-type stars (Holland et al., 1998). Sensitive surveys of 
nearby young stars in the northern hemisphere would help a lot to constrain 
the amount of dust contained in such debris disks. 
\end{itemize}

\subsection{Mapping the CMB anisotropies}
At a wavelength of 2.1~mm, it seems that the measurement of the
Sunyaev-Zeldovich effect is the least affected by radio sources and
dusty galaxies (see the review by Birkinshaw 1998). Mapping the SZ
effect with a comptonisation parameter $y$ sensitivity better than
$1.5\times 10^{-5}\, 1\sigma$ per diffraction beam (20~arcsec FWHM)
in the core of clusters would be possible in only ten hours. This
would be a factor 10-100 increase in mapping speed as compared to
SuZie and Diabolo present bolometer experiments. This might be crucial
for the follow-up of XMM observations (made with a beam of 15~arcsec)
of clusters of galaxies. The other Cosmic Microwave Background (CMB)
anisotropies at small scales that are and will be detected by other
experiments (the Ryle Telescope and the VLA) could receive an
independent confirmation--validation at these clean wavelengths.
Sensitivity is the same as above in $\Delta T/T$ units.
Millimetre interferometers cannot achieve the sensitivity quoted above
for extended sources because of large antennas and the lack of short
spacings.

\section{Instrument definition}
\subsection{Requirements}
So far, the mapping speed improvements came by adding single elements
together. The empirical limit seems to be reached at typically 100
elements. It is limited by the workload (the patience of technicians
and engineers: ask SCUBA people for example) of putting things
together and by the homogeneity of the array. In general the worst
pixels are pulling down the overall sensitivity of the instrument.

Several recent developments in bolometer technology have made
integrated arrays possible. Four projects are in various stage of
completion: BOLOCAM is an East Coast+Caltech project (Mauskopf \& Bock
1998) of 150 integrated 300~mK silicon nitride spider-web pixels with
cones separated by one diffraction size to be put at CSO first and
then on the future 50m (at 1 and 2~mm).  SHARC (and further) is an
operational camera (made by Moseley et al. at NASA--GSFC) of a 24
pixel single line that can be stacked to others in the future, and
that works at a temperature of 300~mK and a wavelength of 450~\micron
at the CSO (Wang et al. 1996). In France, the CEA--LETI--Grenoble (P.
Agnese) is developping a $32 \times 32$ square array as the baseline
for the SPIRE bolometer instrument onboard FIRST (200 to 500\micron).
It uses the Silicon chip making process to make a fully integrated
array.  Another development is with the NbSi thin layers by the
IN2P3--CSNSM--Orsay (L.  Dumoulin).

So if these technologies are available in Europe, what could be the
best use of them in the millimetre domain? In what configuration?  We
argue here that to make full use of the multiplex advantage, the cone
at each pixel must be dropped (as is now planned for SPIRE). A cone
optimises the f/D ratio and hence minimises the pixel size. In case of
lenses, the pixel scale at the focal plane is necessarily larger. A
cone also clearly defines the entrance acceptance angle, effectively
defining the pupil and reducing sidelobes. So, why dropping the cones?
First, the new bolometer technology allows larger pixels without loss
in sensitivity (heat capacity is reduced by making thinner
bolometers). Then, by using a cold pupil common to all pixels, one can
still prevent most of the sidelobes and heatload on the detector.

Moreover, additional problems arise from cones that can be solved by
using an appropriate filled array. The most efficient (straight
instead of parabolic) cones, as in the best known examples (37
bolometer MPIfR and SCUBA), are packed at a spacing of only twice the
diffraction size on the sky, thus mapping at a time, a fraction of 1/4
of the available sky. The sky map must be filled with a drizzle
technics using 16 different positions to have a fair sampling. This is
a likely source of noise, because the map is not fully acquired at the
same time. Another matter of concern is the anomalous refraction which
is known to happen at Pico Veleta and at the JCMT. Even a strong
source has an apparent jitter in front of a detector, giving a
so-called source noise. Calibrations and photometrical measurements
are thus more difficult. When reaching the confusion limit, anomalous
refraction may be a strong limitation. Therefore, it seems that a
cleaner and more efficient solution is to have a filled array of
pixels covering the largest available sky but also at the same time,
fairly sampling the whole available sky (say at half the diffraction
per pixel) in front of a cold pupil. This is the current basic design
of most infrared cameras. The SHARC experiment is already designed
this way.

So far the available arrays are modulated with a wobbling secondary. A
total power readout technique could alleviate the use of a wobbler.
This is already in use by small bolometer arrays: SuZie, NOBA, and
Diabolo at POM2. In the case of a large array, the most promising
observing technique is to fix the telescope in local coordinates ahead
of the target and let the sky drift with the diurnal motion. Local
effects and flat field can thus be disantangled from the real sources.

\subsection{A preliminary implementation}

%______________________________________________________________
%                                                full page figure
%----------------------------------------------------------- 

\setlength\unitlength{1cm}
\begin{figure*}[thp]
  \begin{picture}(15,20)(0,0)
    \includegraphics{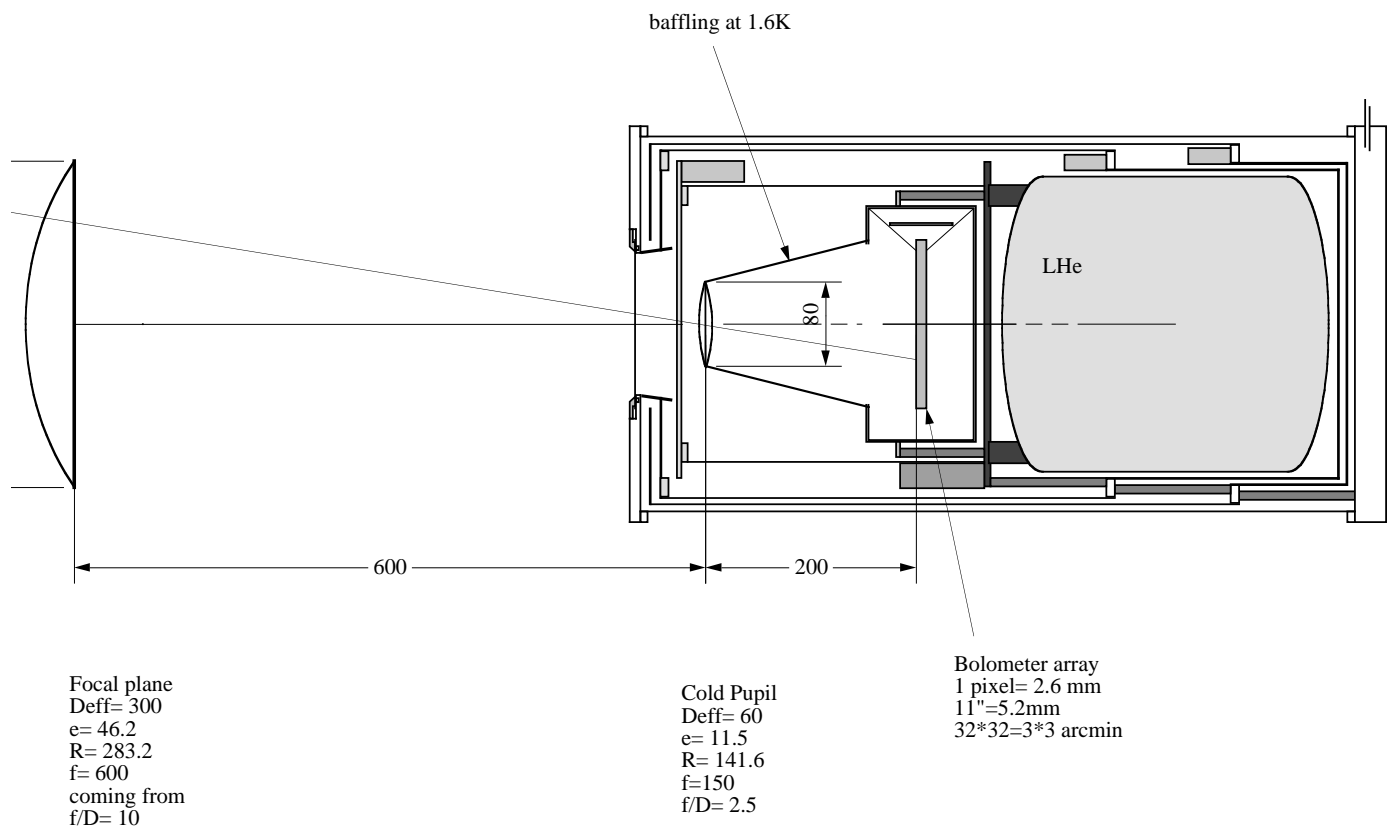}
  \end{picture}
  \caption{A schematic optical layout}
  \label{fi:cao}
\end{figure*}
%______________________________________________________________

Figure~\ref{fi:cao} shows a possible optical layout of the bolometric
camera at 1.2~mm.  It uses one warm lens (assumed here in polyethylene
with a n=1.47 index of refraction) at the 30m focal plane and one cold
lens at the pupil image of the secondary. This cold pupil lens closes
the 1.6K box. Note that the pixel size is here 2.6~mm (i.e. larger
than the operating wavelength) and that is samples half a diffraction
size. Filters (not shown) have to be placed at the cold pupil or just
in front of it at 4~K or higher. The camera at 2.1~mm may require
bigger pixels, hence may be 16 by 16 pixel wide. The field of view
would typically be of 3 by 3 arcminutes, i.e 256 independent beams at
1.2~mm.

Table~\ref{ta:sen} gives the expected sensitivity as conservative
estimates. For that, we assume a very mediocre state of the atmosphere
and the telescope: atmospheric opacity (at the measurement elevation)
and temperature of resp. 0.4 at 1.2~mm and 250 K, telescope emissivity
and main beam efficiency of resp. 0.1 and 0.25 at 1.2~mm and 0.50 at
2.1~mm. The filtering is assumed to have an overall transmission of
15~percent in a $\delta \lambda/\lambda\simeq 0.30$ bandwidth. The box
enclosing the detector must be kept at 1.6K to avoid overloading the
detectors. Most of the photon noise is due to the atmosphere, and not
to the telescope. The same calculation adapted to the present
37-bolometer array and Diabolo experiments give sensitivities which
are slightly above what has been obtained on the sky. The needed
detector sensitivity can be achieved with the present technology, on
single bolometers, especially with relatively slow time constant. This
sensitivity of arrays should be coming soon. Cooling the detector to
0.1~K might be advantageous in this respect.

%______________________________________________________________
\begin{table}
\caption{\label{ta:sen} Sensitivity evaluation}
%%%\begin{flushleft}
\begin{center}
\begin{tabular}{|ll|c|c|} \hline
Characteristics     &  units                          &{\bf 1} &{\bf 2} \\
\hline \hline
Wavelength          & mm                              & 1.2 & 2.1 \\
\hline
Heat load           & pW/pix                          & 23  & 2.5 \\
\hline
Photon noise        & $10^{-17}\zu WHz^{-0.5}$        & 17  & 5 \\
\hline
Assumed Pix. noise  & $10^{-17}\zu WHz^{-0.5}$        & 10  & 5 \\
\hline
Point Source 1$\sigma$, 1s. & mJy                     & 50  & 25 \\ 
\hline
\end{tabular}
\end{center}
%%%\end{flushleft}
\end{table}
%______________________________________________________________

\subsection{A list of potential problems}

We list here several open issues that should be dealt with before
designing such an instrument.

\begin{itemize}
\item Which array of detectors can we foresee to use?

\item Should we use lenses or ellipsoid mirrors as in SHARC? 
  
\item Filtering: use a dichroic to have simultaneously the 1 \& 2mm
  channels (as in Diabolo) or use a filter wheel or make two separate
  cryostats on a similar design to match the detector to the
  wavelength?

\item The readout technique is not yet settled and depends on the used
  array. Multiplexing bolometers has not yet been reported.

\item Cooling to 0.3~K or 0.1~K, with cryocoolers or a cryostat? 

\item Stray light should be a major concern at the start. Ray-tracing
  and Gaussian optics should be used to predict and deal with the
  biggest sources of stray light and ghost images. Warm baffling may
  be used to avoid modulated stray light.

\end{itemize}

\section{Conclusions}

Having a truly mapping millimetre instrument would bring the same
qualitative changes as we saw 15 years ago when IR cameras arrived at
the telescope and replaced single element detectors. The modern
submillimetre instruments are near or at the confusion limit in
extragalactic and galactic environments. Data acquired with arrays
having cones may be very hard to exploit. A true camera has a
potentially large multiplex gain and cleaner behaviour at the
confusion limit. The IRAM 30m user community clearly has to discuss
the various options before attempting to build such an instrument. We
think the challenge is really worth the efforts and that the time is
ripe to start a definition study. We here suggest to continue
pre--design studies and then build the instrument which could be soon fitted with prototypical
detectors of 5 by 5 or 8 by 8  but which would also be compatible
with future 32 by 32 bolometer arrays.

\acknowledgements{ We thank P. Agnese, L. Dumoulin, A. Dutrey, S.
  Guilloteau, J.-M. Lamarre and B. Lazareff for many helpful discussions.}

\end{document}